\documentclass[twocolumn,showpacs,preprintnumbers,superscriptaddress,amsmath,amssymb,floatfix,aps]{revtex4}

\usepackage[dvips]{graphicx}
\usepackage{dcolumn}
\usepackage{bm}

\begin{document}

\title{Plasmons enhance near-field radiative heat transfer for
graphene-covered dielectrics }

\author{V. B. Svetovoy}
\affiliation{MESA$^+$ Institute for Nanotechnology, University of
Twente, PO 217, 7500 AE Enschede, The Netherlands}
\affiliation{Institute of Physics and Technology, Yaroslavl Branch,
Russian Academy of Sciences, 150007, Yaroslavl, Russia}

\author{P. J. van Zwol}
\affiliation{Institut N\'{e}el, CNRS and Universit\'{e} Joseph
Fourier Grenoble, Bo\^{i}te Postale 166, FR-38042 Grenoble Cedex 9,
France}

\author{J. Chevrier}
\affiliation{Institut N\'{e}el, CNRS and Universit\'{e} Joseph
Fourier Grenoble, Bo\^{i}te Postale 166, FR-38042 Grenoble Cedex 9,
France}

\date{\today}

\begin{abstract}

It is shown that a graphene layer on top of a dielectric slab can
dramatically influence the ability of this dielectric for radiative
heat exchange. Effect of graphene is related to thermally excited
plasmons. Frequency of these resonances lies in the terahertz region
and can be tuned by varying the Fermi level through doping or gating.
Heat transfer between two dielectrics covered with graphene can be
larger than that between best known materials and even much larger at
low temperatures. Moreover, high heat transfer can be significantly
modulated by electrical means that opens up new possibilities for
very fast manipulations with the heat flux.

\end{abstract}

\maketitle


Radiative heat transfer (RHT) in vacuum at small distances between
bodies is much increased in the near-field regime as compared to that
given by the black body law \cite{Pol71,Mul01,Kit05}. It happens due
to interaction of evanescent waves at distances small in comparison
with the thermal wavelength $\lambda_T=\hbar c/T$ (here $k_B=1$).
Particularly strong enhancement occurs when bodies can support
surface modes such as plasmon-polaritons and phonon-polaritons
\cite{Jou05,She09}. This effect can be used to improve performance of
near-field photovoltaic devices \cite{Lar06}, in nanofabrication
\cite{Liu05}, and in near-field imaging systems \cite{Wil06}.

Graphene attracted recently enormous attention as a two dimensional
carbon material with unusual electronic properties \cite{Cas09}. It
is considered as a promising material for the development of
high-performance electronic devices \cite{Bol08,Lin11}. Plasmons in
graphene show favorable behavior for applications such as large
confinement, long propagating distances, and high tunability via
electrostatic gating \cite{Jab09}. In contrast with nobel-metals the
plasmon frequencies lie in the terahertz region that is interesting
for radiative heat transfer, but the topic was not explored yet. Heat
transfer was considered \cite{Per10} only between closely spaced
graphene and SiO$_2$ substrate where plasmons do not play significant
role.

Pristine graphene at zero temperature does not support plasmon
excitations but doped material does \cite{Wun06}. On the other hand
at finite temperature plasmons exist even for undoped material
\cite{Vaf06}. These thermoplasmons were shown to change significantly
the thermal Casimir force for suspended graphene \cite{Gom09} and
graphene-covered materials \cite{Sve11b}. In this paper we show that
plasmon excitations in graphene have striking effect on the
near-field RHT between bodies if at least one of them is covered with
graphene.

Usual local materials have fixed frequencies of phonon-polariton or
plasmon-polariton resonances. Graphene is essentially nonlocal
material and its plasmon frequency changes with the wavenumber.
Moreover, it varies significantly with the doping level. These
properties make plasmons in graphene a convenient tool to control the
heat transfer between bodies.

To evaluate the RHT between two bodies 1 and 2 one has to know the
reflection coefficients $r_1^{\mu}$ and $r_2^{\mu}$ for each body as
functions of the frequency $\omega$ and the wave vector
${\textbf{q}}$. These coefficients are different for each
polarization $\mu=s$ (transverse electric) or $\mu=p$ (transverse
magnetic). If the separation $d$ between two parallel plates is small
in comparison with the thermal wavelength, $d\ll \lambda_T$, only
evanescent waves will contribute to the heat transfer. Defining the
heat transfer coefficient (HTC) as $h=\Phi(T,T-\Delta T)/\Delta T$
for $\Delta T\rightarrow 0$, where $\Phi(T_1,T_2)$ is the heat flux
and $\Delta T=T_1-T_2$ is the temperature difference between bodies,
one finds $h=h_p+h_s$, where
\begin{equation}\label{HTC}
    h_{\mu}=\frac{1}{4\pi^2d^2}\int
    \limits_0^{\infty}\textrm{d}\omega\frac{\left(\hbar\omega/T\right)^2
    e^{\hbar\omega/T}}{\left(e^{\hbar\omega/T}-1\right)^2}
    \int\limits_{0}^{\infty}
    \textrm{d}x\frac{xe^{-x}\textrm{Im}r_1^{\mu}\textrm{Im}r_2^{\mu}}
    {\left|1-r_1^{\mu}r_2^{\mu}e^{-x}\right|^2}.
\end{equation}
Here the integration variable in the physical terms is
$x=2d\sqrt{q^2-\omega^2/c^2}$.

Suppose that the body $i$ is a dielectric substrate with the
dielectric function $\varepsilon_i(\omega)$ covered with a layer of
graphene. Reflection coefficient for this body can be found
describing graphene as a sheet with a current. Then for
p-polarization one has \cite{Fal07}:
\begin{equation}\label{rp}
    r_{i}=\frac{k_0\varepsilon_i-k_i+\left(4\pi\sigma/\omega\right)k_0
    k_i}{k_0\varepsilon_i+k_i+\left(4\pi\sigma/\omega\right)k_0
    k_i},
\end{equation}
where $k_0=\sqrt{\omega^2/c^2-q^2}$ and
$k_i=\sqrt{\varepsilon_i\omega^2/c^2-q^2}$ are the  normal components
of the wave vectors in vacuum and in the substrate, respectively, and
$\sigma(q,\omega)$ is the two-dimensional (2D) dynamical conductivity
of graphene. We omitted the polarization index in (\ref{rp}) because
the contribution of graphene can be found using only p-polarization.
Moreover, similar to the situation with the Casimir force
\cite{Gom09,Sve11b} (see also \cite{Fia11}) we can use the
non-retarded approximation to calculate the effect of graphene.
Retardation and the contribution of s-polarization both are
suppressed at least by the factor $v_F/c\approx 1/300$, where $v_F$
is the Fermi velocity in graphene. However, it has to be stressed
that the non-retarded limit can be applied only to the graphene
contribution \cite{Sve11b}. For the HTC this contribution is defined
as
\begin{equation}\label{diff}
    \Delta h=h_p(r_1,r_2)-h_p(r_{10},r_{20}),
\end{equation}
where $h_p$ must be understood as a functional defined by Eq.
(\ref{HTC}) and $r_{i0}$ is the reflection coefficient of the body
$i$ without graphene. To calculate any term $h(r_{1},r_{2})$ or
$h(r_{10},r_{20})$ separately one has to include both the
polarization and retardation effects.

Dielectric function of graphene on the interface of vacuum and body
$i$ can be calculated in the random phase approximation \cite{Cas09}
as $\varepsilon(q,\omega)=1+v_c(q)\Pi(q,\omega)$, where $v_c=2\pi
e^2/\kappa_i q$ is the 2D Coulomb interaction, $\kappa_i$ is defined
by the environment of the graphene layer,
$2\kappa_i=\varepsilon_i(0)+1$, and $\Pi(q,\omega)$ is the 2D
polarizability given by the bare bubble diagram. The latter was
calculated in many papers \cite{Wun06,Kot10}; here we are using the
result \cite{Sve11b} that can be applied for both finite temperature
and finite Fermi level. One can equally express the result via 2D
susceptibility $\chi(q,\omega)=(e^2/q^2)\Pi(q,\omega)$, which, in
turn, is expressed via 2D conductivity as
$-i\omega\chi(q,\omega)=\sigma(q,\omega)$. In this way the reflection
coefficient $r_i$ in the non-retarded limit ($c\rightarrow\infty$)
can be expressed via the dielectric functions of the substrate
$\varepsilon_i(\omega)$ and graphene $\varepsilon(q,\omega)$:
\begin{equation}\label{rp_nr}
    r_{i}=\frac{\varepsilon_i-1+2\kappa_i(\varepsilon-1)}
    {\varepsilon_i+1+2\kappa_i(\varepsilon-1)},\ \ \ c\rightarrow\infty.
\end{equation}

Random phase approximation describes collisionless electron gas, but
this approximation is not sufficient for the RHT. This is because the
graphene contribution given by Eqs. (\ref{diff}) and (\ref{HTC})
disappears in the limit $\textrm{Im}\varepsilon\rightarrow 0$ and
finite dissipation is of principal importance. At low frequencies the
main dissipation in graphene is due to impurities and defects for
electrons \cite{Jab09}. It can be included via the finite relaxation
time $\tau\equiv\gamma^{-1}$ as proposed by Mermin \cite{Mer70},
where $\gamma=10^{12}-10^{13}$ rad/s. In this approach the dielectric
function $\varepsilon_{\tau}(q,\omega)$ is expressed via the
collisionless function $\varepsilon(q,\omega)$ taken at complex
frequency $\omega\rightarrow\omega+i\gamma$ as
\begin{equation}\label{eps_tau}
    \varepsilon_{\tau}(q,\omega)=1+\frac{(\omega+i\gamma)
    \left[\varepsilon(q,\omega+i\gamma)-1\right]}{\omega+i\gamma
    \frac{\varepsilon(q,\omega+i\gamma)-1}
    {\varepsilon(q,0)-1}}.
\end{equation}
Function $\varepsilon_{\tau}$ has to be used in Eq. (\ref{rp_nr})
instead of $\varepsilon$.

At finite doping and finite temperature the function
$\varepsilon(q,\omega)$ can be presented in an analytic form in the
limit $Q=\hbar v_F q/2T\ll 1$ \cite{Sve11b}. In this limit the
intraband transitions dominate in the dielectric function, while the
interband transitions are suppressed by the factor $Q$ and the
dielectric function is
\begin{equation}\label{DED0}
    \varepsilon(q,\omega)=1+\frac{2\alpha_gG(\epsilon_F)}{Q}
    \left(1-\frac{\omega}{\sqrt{\omega^2-v_F^2q^2}}\right),
\end{equation}
where $\epsilon_F=E_F/T$ is the dimensionless Fermi level and
$\alpha_g=e^2/\kappa\hbar v_F$ is the coupling constant in graphene.
Function $G(\epsilon_F)$ is defined as:
\begin{equation}\label{G_def}
    G(\epsilon_F)=\int\limits_{0}^{\infty}\textrm{d}tt
    \frac{1+\cosh t\cosh\epsilon_F}
    {\left(\cosh t+\cosh\epsilon_F\right)^2}.
\end{equation}
For small and large values of $\epsilon_F$ it takes the asymptotic
values $G(\epsilon_F\rightarrow 0)=2\ln2$ and
$G(\epsilon_F\rightarrow\infty )=\epsilon_F$, respectively.

Let us assume first that the substrate of the body 1 has no optical
activity in the terahertz range. It means that
$\varepsilon_1(\omega)\approx \varepsilon_1(0)$. In this case the
reflection coefficient has a pole when $\varepsilon(q,\omega)=0$
($\gamma\rightarrow 0$). This pole describes plasmon in graphene with
the dispersion relation
\begin{equation}\label{omp}
    \hbar\omega_p(q)\approx \sqrt{2\alpha_gG(\epsilon_F)
    \hbar v_FqT},\ \ \ Q\ll 1.
\end{equation}
When $\epsilon_F\rightarrow 0$ we reproduce Vafek's thermoplasmon
\cite{Vaf06}. In the limit of large $\epsilon_F$ the plasmon
frequency does not depend anymore on $T$ and we find the plasmon that
emerging at finite doping \cite{Wun06}.

For the best heat transfer the resonances in the opposing bodies have
to match each other \cite{Jou05}. When one body is covered with
graphene one can always find a value of the wavenumber $q$ that gives
the plasmon resonance matching the resonance in the opposing
body.This simple principle gives qualitative explanations for rich
physics that can be realized between bodies covered with graphene.
Let us illustrate this statement.

\begin{figure}[ptb]
\begin{center}
\includegraphics[width=0.48\textwidth]{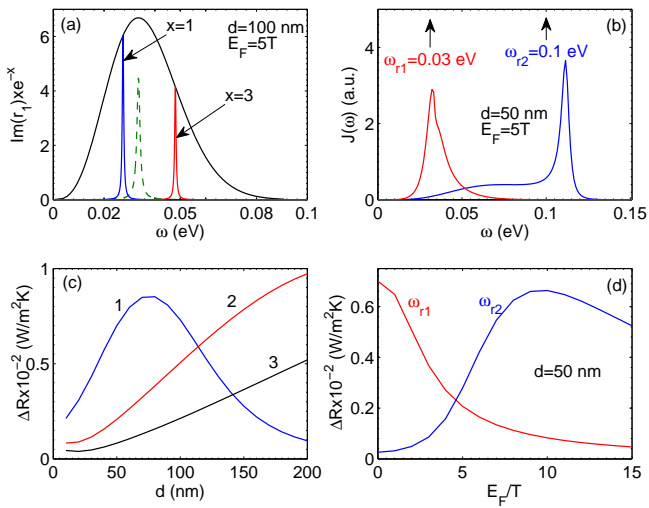}
\vspace{-0.5cm} \caption{(Color online) Inactive dielectric covered
with graphene (body 1) against a dielectric without graphene described by the
Lorentz model (body 2).  (a) Resonances in the reflection
coefficients. The dashed line (green) shows the fixed resonance
in $\textrm{Im}r_2$. Plasmon resonances in the body 1 are shown for
$x=1$ and $x=3$ together with the enveloping line (black). (b) Spectral
density for two values of the Lorentz resonance. (c) $\Delta R$ as a
function of distance. Curves 1, 2, 3 correspond to $\epsilon_F=$0, 5,
10, respectively. (d) $\Delta R$ as a function of the Fermi level.  }
\label{fig1}
\vspace{-0.9cm}
\end{center}
\end{figure}

Suppose that the second body can be described at low frequencies by a
single Lorentz-Drude oscillator with the dielectric function
\begin{equation}\label{LO}
    \varepsilon_{2}(\omega)=\varepsilon_{\infty}\left(1+
    \frac{A\omega_r^2}{\omega_r^2-\omega^2-i\Gamma\omega}\right),
\end{equation}
where $\omega_r$ is the resonance frequency, $\Gamma\ll\omega_r$ is
the resonance width, $A$ is the amplitude, and $\varepsilon_{\infty}$
is the high-frequency dielectric constant. Reflection coefficient
$r_2$ has the resonance at frequency corresponding to the surface
wave excitation that is determined by the equation
$\varepsilon_2(\omega)=-1$. This frequency is
\begin{equation}\label{om2}
    \omega_{2s}=\omega_r\sqrt{B},\ \ B=\frac{\varepsilon_2(0)+1}
    {\varepsilon_{\infty}+1}.
\end{equation}
Imaginary part of $r_2$ is shown in Fig. \ref{fig1}(a) by the dashed
line. Plasmon resonances in graphene give peaks in $\textrm{Im}r_1$
that are shown for $x=1$ and $x=3$. Because $x=2dq$ there is a value
of $q$ when frequencies of the resonances in both bodies match each
other.


Integration over $x$ in Eq. (\ref{HTC}) gives the spectral density of
the HTC shown in Fig. \ref{fig1}(b) for two values of the Lorentz
resonance. This density is concentrated around the resonance in the
body 2. Integrating over $\omega$ one finds the effect of graphene in
the HTC. In the rest of the paper we present the results for the
scaled HTC defined as
\begin{equation}\label{SHTC}
    \Delta R=(d/100\;\textrm{nm})^2\cdot\Delta h.
\end{equation}
In Fig. \ref{fig1}(c) $\Delta R$ is shown for the Lorentz resonance
$\omega_{r1}=0.03$ eV and three different values of the Fermi level.
We can compare $h=\Delta R\cdot (100\;\textrm{nm}/d)^2$ with the
black body coefficient $h_{bb}\approx 6$ W/m$^2$K. One can see that
graphene provides a significant contribution to the HTC. Dependence
of $\Delta R$ on the Fermi level is nontrivial and strong. It is
shown in Fig. \ref{fig1}(d). Value of $\Delta R$ varies more than 10
times when the Fermi level changes from 0 to $10T$. Moreover, this
dependence changes from decreasing to increasing when $\omega_{r}$
changes from 0.03 eV to 0.1 eV. Indeed, the characteristic plasmon
frequency $\omega_p^{ch}$ is given by Eq. (\ref{omp}) for $q\sim
1/2d$. If $\omega_p^{ch}>\omega_r$ for all $E_F$ then $\Delta R$ will
decrease when the Fermi level increases. In the opposite case $\Delta
R$ will increase with $E_F$ while the condition
$\omega_p^{ch}<\omega_r$ holds true and change to decreasing when the
condition is broken.

\begin{figure}[ptb]
\begin{center}
\includegraphics[width=0.48\textwidth]{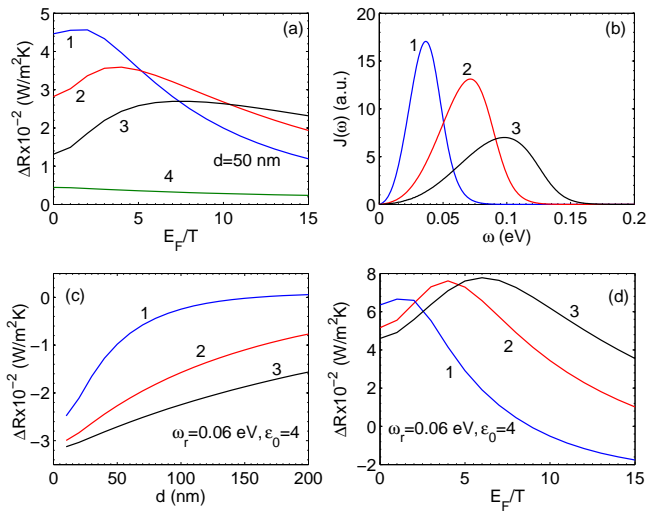}
\vspace{-0.5cm} \caption{(Color online) (a) $\Delta R$ as a function of
$\epsilon_F$ for inactive dielectrics both covered with graphene. Curves
1, 2, 3 are for $\varepsilon_0=$4, 7, 12. Curve 4 is
for $\varepsilon_1(0)=4$ and $\varepsilon_2(0)=7$. (b) Spectral density
for $\varepsilon_0=4$ and $\epsilon_F=$0(1), 5(2), 10(3). (c) Two identical
substrates described with the Lorentz model; one is covered with
graphene; $\epsilon_F=$0(1), 5(2), 10(3). (d) Substrates are the same as
in (c) but both covered with graphene. Dependence on $\epsilon_F$
is shown for $d=50$(1), 100(2), 150(3) nm.} \label{fig2} \vspace{-0.9cm}
\end{center}
\end{figure}

Effect of graphene is especially strong for identical inactive bodies
covered with graphene (see Fig. \ref{fig2}(a)). This is because the
plasmons frequencies in both bodies coincide for every $q$ so that
the spectral density becomes much wider than in the case of one
graphene-covered body. This density is shown in Fig. \ref{fig2}(b)
for different values of $E_F$. The amplitude and width of the curves
is essentially controlled by the thermal factor in Eq. (\ref{HTC}).
Value of $\Delta R$ becomes smaller when the substrate dielectric
constant $\varepsilon_1(0)=\varepsilon_2(0)=\varepsilon_0$ increases.
It is also reduced very fast when the substrates are not the same.
This is because the plasmon frequencies do not match any more due to
different coupling constants $\alpha_g$ in the graphene layers of
different bodies. When one or both of the substrates are metals the
effect of graphene disappears, $\Delta R \rightarrow 0$. It happens
because in the limit $\varepsilon_i\rightarrow\infty$ the reflection
coefficients with and without graphene coincide, $r_i\rightarrow
r_{i0}$.

Suppose that body 1 is covered with graphene but both substrates are
optically active and can be presented by the same Lorentz oscillator.
In this case the body 1 supports two surface waves with the
frequencies
\begin{equation}\label{sw}
    2(\omega_{1s}^2)_{1,2}=\omega_{2s}^2+B\omega_p^2\pm
    \sqrt{(\omega_{2s}^2+B\omega_p^2)^2-4\omega_{2s}^2\omega_p^2}.
\end{equation}
These frequencies do not coincide with $\omega_{2s}$. Only in the
limit $\omega_p\rightarrow 0$ one has $(\omega_{1s})_1\rightarrow
\omega_{2s}$. It means that the resonances in different bodies never
match each other and the term $h_p(r_1,r_2)$ in Eq. (\ref{diff})
cannot be large. On the other hand, the term $h_p(r_{10},r_{20})$
must be large because without graphene on body 1 the resonances
coincide. Therefore, for identical optically active substrates,
graphene on one of the bodies will reduce the heat transfer, $\Delta
R<0$, as demonstrated in Fig. \ref{fig2}(c). Indeed, the total HTC
given by Eq. (\ref{HTC}) is always positive.

When both active substrates are covered with graphene then the
resonances in different bodies match and the spectral density is wide
similar to that shown in Fig. \ref{fig2}(b). Difference with the case
of inactive substrates is that the term $h_p(r_{10},r_{20})$ in Eq.
(\ref{diff}) is nonzero and gives important negative contribution.
Now $\Delta R$ can be negative or positive depending on the
parameters as shown in Fig. \ref{fig2}(d).

For applications in nanoelectronics it is interesting to have a
device that can open and close a heat transfer channel at high
switching rate \cite{Zwo11}. Bodies covered with graphene allow deep
modulation of the heat flux at very high frequencies. The switching
frequency $f$ must be smaller than important frequencies in the
graphene dielectric functions, $f\ll v_F/4\pi d\sim 1$ THz but actual
restriction comes from graphene electronics that can be switched at
frequencies up to 100 GHz \cite{Lin10}. The largest heat flux one can
get for identical substrates inactive in THz range that have low
dielectric constant in this range. Many polymeric materials used in
electronics \cite{Zha10} meet these conditions, for example,
polyimide.

\begin{figure}[t]
\begin{center}
\includegraphics[width=0.48\textwidth]{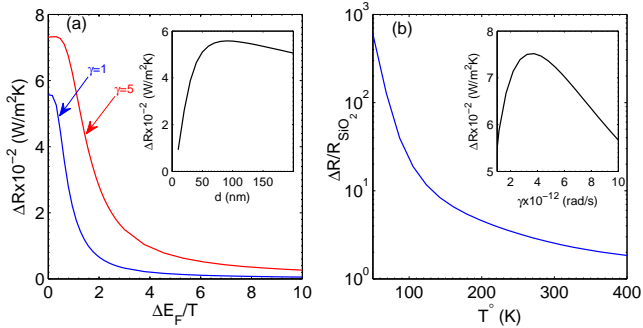}
\vspace{-0.5cm} \caption{(Color online) (a) $\Delta R$ between two
inactive dielectrics ($\varepsilon_0=3$) covered with graphene as a
function of the Fermi level difference $\Delta E_{F}=E_{F1}-E_{F2}$;
$\gamma$ is in units $10^{12}$ rad/s.
The inset shows how the maximal value depends of the
distance $d$. (b) Ratio of the HTC for graphene-covered
dielectrics ($\Delta R$) and two SiO$_2$ plates ($R_{SiO_2}$).The inset
presents dependence of the maximal $\Delta R$ on the relaxation frequency.
} \label{fig3} \vspace{-0.9cm}
\end{center}
\end{figure}

When graphene layers on both substrates are equally doped, then the
HTC is large but does not change much with the Fermi level as shown
in Fig. \ref{fig2}(a). However, when the doping in each layer is
different the HTC decreases rapidly because of mismatch of the
plasmon frequencies in the graphene layers. Due to this dependence
one can modulate the heat transfer between bodies by changing the
difference between the Fermi levels in different bodies. Situation is
illustrated in Fig. \ref{fig3}(a). HTC decreases 30 times or more
when $\Delta E_F$ changes from zero to $10T$. The latter value is
realized for a reasonable difference in the carrier density around
$\Delta n=5\cdot 10^{12}$ cm$^{-2}$ at $T=300^{\circ}$ K. Maximal
value of $\Delta R$ varies to some degree with the relaxation
frequency $\gamma$ in graphene as shown in the right inset and with
the distance $d$ as shown in the left inset. It has to be stressed
that $\Delta R$ between identical inactive dielectrics covered with
graphene is larger than that between two SiO$_2$ plates. Note that
SiO$_2$ is considered as one of the very best material for the
radiative heat transfer, for which $R_{SiO_2}=296$ W/m$^2$K at
$T=300^{\circ}$ K. With temperature decrease $R_{SiO_2}$ decreases
fast due to the thermal factor in (\ref{HTC}), but $\Delta R$
decreases much slower because of plasmon tuning. As the result the
ratio $\Delta R/R_{SiO_2}$ increases significantly with $T$ decrease
as shown in Fig. \ref{fig3}(b).

In conclusion, we analyzed the change in the near-field radiative
heat transfer for materials covered with graphene. Plasmon
excitations in graphene change drastically material's ability to the
heat transfer. Plasmon frequency is tunable and depends on the
wavenumber, Fermi level, temperature, and the substrate dielectric
constant. Plasmons can be tuned with the resonances in the opposite
body to maximize the HTC. The strongest effect is reached for
terahertz-inactive dielectrics with low dielectric constant covered
with graphene. In this case the HTC is larger than for the best known
materials especially at low temperatures.  HTC can be reduced 100
times or so by changing the relative carrier concentration in
different bodies. This opens up the possibility to control the heat
flux at frequencies as high as 100 GHz.


\end{document}